\documentclass[
aps,%
12pt,%
final,%
notitlepage,%
oneside,%
onecolumn,%
nobibnotes,%
nofootinbib,%
superscriptaddress,%
noshowpacs,%
centertags]%
{revtex4}

\usepackage{epsf}

\usepackage{amsmath}
\usepackage{graphicx}
\usepackage[cp866]{inputenc}
\usepackage[english,russian]{babel}

\newcommand{\lan}{\langle}
\newcommand{\ran}{\rangle}

\newcommand{\be}{\begin{equation}}
\newcommand{\ee}{\end{equation}}
\newcommand{\ba}{\begin{aligned}}
\newcommand{\ea}{\end{aligned}}

\newcommand{\bea}{\begin{equation}\begin{aligned}}
\newcommand{\eea}{\end{aligned}\end{equation}}

\newcommand{\beq}{\begin{eqnarray}}
 \newcommand{\eeq}{\end{eqnarray}}

\def\fun#1#2{\lower3.6pt\vbox{\baselineskip0pt\lineskip.9pt
\ialign{$\mathsurround=0pt#1\hfil##\hfil$\crcr#2\crcr\sim\crcr}}}

\newcommand{{\PBC}}{{\rm PBC}}

\begin{document}


\title{Dilaton and gluon condensate in a nucleon medium \\ }
\author{N.O. Agasian}
\email{agasian@itep.ru}
\affiliation{
Institute of Theoretical and Experimental Physics, \\
117218, Moscow, Russia}


\begin{abstract}
The gluon condensate as a function of temperature and baryon density in a nucleon medium
is obtained from an effective dilaton Lagrangian. It is shown that, at a normal nuclear density of nucleons,
 $n_0 = 0.17$ fm$^{-3}$ the gluon condensate decreases by about $10\%$.
\end{abstract}

\maketitle



\vspace{1cm}

It is well known that, at temperatures and baryon
densities below the phase transition point, QCD
is essentially nonperturbative and is characterized
by the phenomena of confinement and spontaneous
chiral symmetry breaking. This is due to the presence
of strong gluon fields in the QCD vacuum, which
make a finite contribution to the shift of the vacuum energy
density through the anomaly in the trace of
the energy-momentum tensor and, thus, through
the gluon condensate $\lan G^2\ran \equiv \lan (gG^a_{\mu\nu})^2\ran $. It is highly
desirable to study the gluon condensate as a function
of the temperature $T$ and the baryon density $n$. Knowing the dependence of $\lan G^2\ran$
on $T$  and  $n$, one
can predict the behavior of physical quantities like the
hadron masses and the constant $f_\pi$  versus medium
features $T$  and $n$. This may be of paramount importance
in studying phenomena in nuclear matter under
anomalous conditions, for example, in heavy-ion
collisions. The dependence of
$\lan G^2\ran$  on $T$  and $n$ was
analyzed in a number of studies (see, for example,
\cite{Cohen:1991nk, Agasian:1993fn, Sollfrank:1994du, Carter:1996rf,
Agasian:1997zr,  Carter:1998ti, Agasian:1999id,
Schaefer:2001cn, Sannino:2002wb, Drago:2001gd,
Agasian:2001bj, Agasian:2003ux})
on the basis of various physical approaches. In
the present study, we employ the low-energy effective
dilaton Lagrangian to explore the dependence of $\lan G^2\ran$ on the temperature and baryon density.
 In contrast to
other approaches, the approach based on the effective
dilaton Lagrangian makes it possible to take into
account the inverse effect of the change in the gluon
condensate on particle masses and to find thereby $\lan G^2\ran$  over the entire region of $T$  and $n$
up to the phase-transition point.

In this study, we formulate a general approach
and find expressions for $\lan G^2\ran$   that are linear in the
particle density, this enabling us to compare the results
produced by the developed method with other known results for
 $\lan G^2\ran$  as a function of temperature
and baryon density.

In gluodynamics, the glueball having the vacuum
quantum numbers $J^{PC} =0^{++}$ is the lightest hadron,
which is therefore stable.
A low-energy Lagrangian
describing the interaction of the $0^{++}$ glueball (dilaton)
and realizing scale-invariance Ward identities, in
just the same way as the chiral pion Lagrangian realizes
chiral Ward identities at the tree level, was constructed
in \cite{Schechter:1980ak, Migdal:1982jp}.
The effective dilaton Lagrangian has the form
 \be
 L(\sigma)=\frac12 (\partial_\mu\sigma)^2-V(\sigma),~~ V(\sigma)
 =\frac{\lambda}{4} \sigma^4 \left (\ln \frac{\sigma}{\sigma_0}
 -\frac14\right).
 \label{1}
 \ee
The field  $\sigma$ is related to the trace of the energy-momentum tensor
$\theta_{\mu\mu}(x)$ in gluodynamics by the equation
\be
\frac{m_0^4}{64|\varepsilon_v|}\sigma^4 (x) =-\theta_{\mu\mu}
(x) =\frac{b}{32\pi^2} (gG^a_{\mu\nu} (x))^2.
\label{2}
\ee
Here, $b$ is a coefficient in the Gell-Mann-Low function,
while the constants  $\lambda$ and $\sigma_0$ are expressed in terms of physical parameters as
\be
\lambda=\frac{m^4_0}{16|\varepsilon_v|},~~ \sigma^2_0 =
\frac{16|\varepsilon_v|}{m^2_0},
\label{3}
\ee
where  $|\varepsilon_v| =-\frac14 \lan \theta_{\mu\mu}\ran$
is the nonperturbative vacuumenergy
density and $m_0$ is the dilaton mass.

As was indicated above, the effective dilaton Lagrangian
can be used to find the gluon condensate
at finite density and temperature. We now investigate
the behavior of  $\lan G^2\ran$ in a nuclear medium.We address
the question of how the gluon vacuum changes in the
case where the medium is a perfect Fermi gas of nucleons.
It is of crucial importance to understand first
how the Fermi filling of nucleons deforms the gluon
condensate, whereupon one can take into account
the effects of the interaction between particles of the medium.

From low-energy QCD theorems \cite{Migdal:1982jp, Novikov:1981xj} and
from scale invariance, it follows that nucleon interaction
with the dilaton is described by the Lagrangian
 \be
 L_{\sigma NN} = m^*_N\bar \psi \psi,~~
 m^*_N=m_N\frac{\sigma}{\sigma_0}
 \label{4}
 \ee
In order to determine the dependence of  $\lan G^2\ran$ on
 $T$  and $n$, we write a thermodynamic potential for nucleons
interacting with the dilaton. In general, we have
$$ \Omega [\sigma,\mu_p,\mu_n] =V(\sigma) -V(\sigma_0)
-\frac{T}{V} \ln Z [\sigma, \mu_p, \mu_n],$$
\be
Z=\int\prod_{i=p,n} D\bar \psi_i D\psi_i \exp \left\{ -
\int^{1/T}_0 d\tau \int_V d^3 x \sum_{i=p,n} [\bar \psi_i (\hat
p-m^*_N) \psi_i +\mu_i\psi^+_i \psi_i]\right\}
\label{5}
\ee
where $\mu_p$ and $\mu_n$ are the chemical potentials of protons
and neutrons, respectively.

Let us consider an isotopically invariant medium, $\mu_p=\mu_n=\mu$.
The thermodynamic potential for nucleons can then be represented in the form
\be
\Omega_N = 4 T\int\frac{d^3p}{(2\pi)^3} \ln \left( 1+
e^{(\mu-\sqrt{p^2+m^{*2}_N})/T}\right ) + 4
T\int\frac{d^3p}{(2\pi)^3} \ln \left( 1+
e^{-\sqrt{p^2+m^{*2}_N}/T}\right )
\label{6}
\ee
Here, the second term describes the contribution to
the thermodynamic potential for antinucleons.

Minimizing the expression $ \Omega [\sigma,\mu_p,\mu_n]$  with respect
to the field $\varphi={\sigma}/{\sigma_0}$, we find the equilibrium
value of the dilaton field at given $T$ and $\mu$. The result is
\be
4 |\varepsilon_v| \varphi^3 \ln \varphi -  m^2_N \varphi [
F(T,\mu) + F(T, \mu=0) ] =0,
\label{7}
\ee
where
 \be F(T,\mu)
=\int\frac{d^3p}{(2\pi)^3} \frac{1}{\sqrt{p^2+ m_{N}^{*2}}
(e^{(\sqrt{p^2+ m_{N}^{*2}}-\mu)/T}+1)}.
\label{8}
\ee
The dependence of the gluon condensate on temperature
and the chemical potential is determined by the relation
\be
\frac{\lan
G^2\ran_{T,\mu}}{\lan G^2\ran} \equiv \varphi^4,
\label{9}
\ee
where $\varphi$ is a solution to Eqs. (\ref{7}) and (\ref{8}).

Let us consider the case of $\mu=0$, that is, the case
of a hot nucleon-antinucleon gas.
At temperatures
satisfying the condition $T\ll m_N$ we can set $ m^*_N=m_N$, whereupon
Eq.  (\ref{7})  assumes the form
\be
2 |\varepsilon_v|
\varphi^2 \ln \varphi -  m^2_N F[T,\mu=0]=0
\label{10}
\ee
In the nonrelativistic approximation  $(T\ll m_N)$, we have
\be F(T,\mu=0)
=\frac{Tm}{2\pi^2 } K_1 \left( \frac{m_N}{T}\right) \simeq
\frac{T^{3/2} m^{3/2}_N}{(2\pi)^{3/2}} e^{-m_N/T}.
\label{11}
\ee
Equation  (\ref{10}) then reduces to the form
\be \varphi^2 \ln
\varphi - \frac{1}{2|\varepsilon_v|} \frac{T^{3/2}
m_N^{5/2}}{(2\pi)^{3/2}} e^{-m_N/T} =0.
\label{12}
\ee
This equation has the following solution:
 \be
 \varphi=1
-\frac{1}{2|\varepsilon_v|}\frac{T^{3/2} m_N^{5/2}}{(2\pi)^{3/2}}
e^{-m_N/T}+ O(e^{-2m_N/T}).
\label{13}
\ee
Using relation  (\ref{9}), we accordingly find for the gluon condensate that
\be
\lan G^2\ran_T =\lan G^2\ran \left(1
-\frac{2}{|\varepsilon_v|}\frac{T^{3/2} m_N^{5/2}}{(2\pi)^{3/2}}
e^{-m_N/T}\right)
\label{14}
\ee

Let us consider the case of zero temperature and a
finite baryon density. Using standard thermodynamic
relations and going over from the chemical potential
to the baryon density and from the thermodynamic
potential to the energy density, we arrive at
\be
\varepsilon [\sigma_n,n]=V(\sigma_n) - V(\sigma_0) + 4
\int^{p_F}_0 \frac{d^3p}{(2\pi)^3}
\sqrt{p^2+{m^*_N}^2}.
\label{15}
\ee
Here, $n=n_n+n_p=2p^3_F/3\pi^2$ is the baryon density
and the chemical potential is related to
 $p_F$  by the standard equation $\mu^2=p^2_F+{m^*_N}^2$. The field $\sigma_n$
is a function of the baryon density, while $\sigma_0$ is the dilaton field
at  $n=0$.
At densities of about the nuclear density,
$ n=n_0 = 0.17$ fm$^{-3}$, we have
 $p^2_F\ll m^2_N$; therefore, we can use the nonrelativistic approximation,
  \be
  \sqrt{p^2+{m^*_N}^2} = m^*_N+\frac{p^2}{ 2m^*_N},~~ m^*_N=m_N
  \frac{\sigma_n}{\sigma_0}
  \label{16}
  \ee
Substituting (\ref{16}) into (\ref{15}) and retaining the lowest order terms in the density, we obtain
  \be
  \varepsilon [\varphi, n] =V(\varphi) - V(1) + m_N n\varphi + O(n^2).
  \label{17}
  \ee
The equation for the equilibrium value of the dilaton
field $\varphi= \sigma_n/ \sigma_0$ in a medium is found by minimizing
the energy density. Varying expression (\ref{17}) with respect
to   $\varphi$, we arrive at
  \be
  16 |\varepsilon_v| \varphi^3 \ln \varphi + m_N n=0
  \label{18}
  \ee
  In the case of $ m_Nn\ll 16 |\varepsilon_v|$, Eq. (\ref{18}) has a solution
of the form
    \be
    \varphi=1-m_Nn/ 16|\varepsilon_v|
    \label{19}
    \ee
Thus, we finally obtain the baryon-density dependence
of the gluon condensate in the form \cite{Cohen:1991nk, Agasian:1999id}

  \be
  \lan G^2\ran_n =\lan G^2\ran (1-
  m_Nn/4|\varepsilon_v|)
  \label{20}
  \ee

In QCD, $|\varepsilon_v| \simeq 3.375$ GeV$^4$ \cite{Shifman:1978bx} ,
and we obtain $ m_Nn/4 |\varepsilon_v|\simeq 0.1$ at the normal nuclear density $n=n_0$.
We can therefore see that, at densities on this
order of magnitude, the gluon condensate decreases
by a value of about $\sim 10\%$.

I am grateful to Yu.A. Simonov for stimulating discussions.

This work was supported by a federal program
(no. 40.052.1.1.1112) of the Ministry of Industry, Science,
and Technologies of Russian Federation and was also
funded by a grant (no. 3855.2010.2) for support of leading scientific schools.


\end{document}